%% file: main.tex
\documentclass[10pt]{article}

\usepackage[a4paper, inner=1.5cm, top=1.5cm, outer=1.5cm, bottom=1.5cm, includefoot]{geometry}
\usepackage{amsmath}
\usepackage{bm}
\usepackage[max2]{authblk}
\usepackage{amssymb}
\usepackage{mathrsfs}
\usepackage{graphicx}
\usepackage{amssymb}
\usepackage{wrapfig}
\usepackage{enumitem}
\usepackage{tikz-network} 
\usepackage{tabularx}
\usepackage[ruled,lined,algosection]{algorithm2e}
\usepackage[
  natbib=true,
  url=false,
  isbn=false,
  mincitenames=1,
  maxcitenames=2,
  sorting=none,
  sortcites=true,
  backend=biber
  ]{biblatex}
  
\usepackage{amsthm}
\usepackage{appendix}
\usepackage{subdepth}
\usepackage{float}
\usepackage{hyperref}
\hypersetup{
    colorlinks,
    linkcolor={red!50!black},
    citecolor={blue!50!black},
    urlcolor={blue!80!black}
}
\usepackage{mathtools}
\usepackage{multirow}
\usepackage{array}
\usepackage{mathtools}
\usepackage{diagbox}
\usepackage{placeins}
\usepackage{abstract}
\usepackage{listings}
\usepackage[T1]{fontenc}
\usepackage{minted}
\usepackage{xcolor} 
\usepackage{afterpage} 
\usepackage{pdflscape} 

\DeclareUnicodeCharacter{03B3}{$\gamma$}
\DeclareUnicodeCharacter{2212}{--}
\DeclareUnicodeCharacter{03B1}{$\alpha$}

\DeclareFieldFormat{date}{\thefield{year}}

\input{style}

\lstdefinelanguage{yaml}{
  keywords={study, job, run},
  sensitive=false,
  comment=[l]{\#},
  moredelim=[l][\color{gray}\bfseries]{:}  
}

\title{\textbf{AutoPot}: \textbf{Auto}mated and massively parallelized construction of  Machine-Learning \textbf{Pot}entials}
\author[1]{M. Hodapp\thanks{maxludwig.hodapp@mcl.at}}
\author[2]{G. Anciaux\thanks{guillaume.anciaux@epfl.ch}}
\affil[1]{Christian Doppler Laboratory for Digital material design guidelines for mitigation of alloy embrittlement, Materials Center Leoben Forschung GmbH (MCL), Leoben (AT)}
\affil[2]{Ecole Polytechnique Federale de Lausanne (EPFL), Lausanne (CH)}

\renewcommand\footnotemark{}

\begin{document}



\maketitle

\begin{abstract}
    Machine-learning potentials (MLIPs) have been a breakthrough for computational physics in bringing the accuracy of quantum mechanics to atomistic modeling.
    To achieve near-quantum accuracy, it is necessary that neighborhoods contained in the training set are rather close to the ones encountered during a simulation.
    Yet, constructing a single training set that works well for all applications is, and likely will remain, infeasible, so, one strategy is to supplement training protocols for MLIPs with additional learning methods, such as active learning, or fine-tuning.
    This strategy, however, yields very complex training protocols that are difficult to implement efficiently, and cumbersome to interpret, analyze, and reproduce.
    
    To address the above difficulties, we propose AutoPot, a software for automating the construction and archiving of MLIPs.
    AutoPot is based on BlackDynamite, a software that operates parametric tasks, e.g., running simulations, or single-point ab initio calculations, in a highly-parallelized fashion, and Motoko, an event-based workflow manager for orchestrating interactions between the tasks.
    The initial version of AutoPot supports selection of training configurations from large training candidate sets, and on-the-fly selection from molecular dynamics simulations, using Moment Tensor Potentials as implemented in MLIP-2, and single-point calculations of the selected training configurations using VASP.
    Another strength of AutoPot is its flexibility:
    BlackDynamite tasks and orchestrators are Python functions to which own existing code can be easily added and manipulated without writing complex parsers.
    Therefore, it will be straightforward to add other MLIP and ab initio codes, and manipulate the Motoko orchestrators to implement other training protocols.
\end{abstract}




\section{Introduction}

Machine-learning interatomic potentials (MLIPs) \cite{behler_generalized_2007,bartok_gaussian_2010,thompson_spectral_2015,shapeev_moment_2016,zhang_deep_2018,drautz_atomic_2019,pun_physically_2019,unke_machine_2021,batzner_e3-equivariant_2022,batatia_mace_2022,fan_gpumd_2022,hodapp_equivariant_2024,wang_en-equivariant_2024} have revolutionized computational materials science by overcoming the two persisting limitations of empirical interatomic potentials:
\begin{itemize}
    \item
    MLIPs are able to predict mechanisms that influence defect motion, stability of crystal structures, etc., of a hypothetical large-scale quantum-mechanical simulation, enabling quantum-accurate multiscale simulations up to the continuum level (e.g., \cite{mortazavi_recent_2025}).
    \item
    MLIPs are able to predict the influence of variations of the chemistry on material properties that, e.g., enables exploring trends in the composition of complex concentrated alloys (e.g., \cite{byggmastar_modeling_2021,moitzi_ab_2024,song_general-purpose_2024}).
\end{itemize}

Despite these success stories, achieving predictive accuracy requires a carefully designed training set containing atomic neighborhoods that are not too far from those encountered during a simulation of the material behavior, which is a time-consuming task.
Random sampling of training configuration before running a simulation is often not accurate enough, or requires a lot of training configurations \cite{zhang_atomistic_2023,ito_machine_2024,lin_machine-learning_2024,liu_learning_2024,erhard_how_2025,shuang_modeling_2025}, so, protocols for constructing MLIP training sets are complemented with advanced sampling algorithms, e.g., active learning \cite{settles_active_2009} from simulations that resemble the underlying problem of interest \cite{podryabinkin_active_2017,zhang_active_2019,vandermause_--fly_2020,hodapp_operando_2020,podryabinkin_nanohardness_2022}.
Another related problem is fine-tuning a MLIP that has been pre-trained on a (generally very large) dataset using additional problem-specific configurations (e.g., \cite{deng_overcoming_2024}).

While the above algorithms have been successfully applied to construct reliable MLIPs, it remains challenging to implement them efficiently.
Another challenge is the complexity of the algorithms requiring communication between several software codes, i.e., the code that implements the potential, the code that runs the ab initio calculations, and a code that runs the simulation from which configurations are sampled.
Developing and implementing such algorithms becomes more and more difficult with an increasing complexity of MLIPs.
For example, \citet{kotykhov_actively_2025,burov_active_2026} proposed active learning algorithms that train magnetic MLIPs on spin-constrained Density Functional Theory (DFT) calculations during MD simulations.
Capturing realistic atomic trajectories of magnetic materials requires a hybrid approach combining molecular dynamics and Monte Carlo simulations, so, potentially two simulation codes from which configurations are sampled.
Hence, making such flexible algorithms constructing MLIPs more reliable and reproducible requires streamlining training protocols.

At the center of active learning stands the creation and storage of
configuration datasets, a \textit{sine qua non} condition to incrementally train
MLIPs. New configurations will be dynamically selected, created, and
interleaved, with training phases. For reproducibility reasons, it is crucial to
keep the link existing between intermediate results down to the final potential.
Therefore, the storage of configurations, training sets, algorithms inputs and
outputs (MD, DFT, training), need to be saved and organized to enable a machine
learning model following open science and FAIR \cite{noauthor_fair_nodate}
principles.

With similar goals, there exist several ML-lifecycle trackers, for general
purposes that keep track of hyper-parameters, datasets, versioning, and model
access. Training sets are usually preserved in file stores, with associated
metadata (e.g., dataset/model unique identifiers). Executing a workflow
description -- \textit{what should run, in what order, where, and with what
data} -- is the role of a workflow execution platform, i.e. scheduling and
running tasks on HPC/cloud resources, retrying failed jobs, and tracking the
execution graph. General-purpose executors\cite{noauthor_snakemake,
noauthor_flyte_nodate, noauthor_fireworks, noauthor_jobflow-remote,
noauthor_mlops_nodate, noauthor_blackdynamite_2024, noauthor_bamboost_nodate}
usually offer Python for workflow descriptions, enabling necessary
interoperability for FAIR\cite{noauthor_fair_nodate} compliance. Execution graph
metadata can be preserved in relational databases (SQL-like), object databases
(ZEO), or as structured files (JSON, MongoDB). Larger artifacts are usually
stored on filesystems or object storage repositories (Amazon-S3).

Furthermore, the above tools have generally a destructive approach regarding
intermediate results, which may prevent full reproducibility. Concretely, a
workflow manager such as \textit{Flyte} \cite{noauthor_flyte_nodate} will use
anonymous runs and preserve only minimal (meta-)data in output databases. This
cloud-based relation with data storage is frugal, but requires the workflow to
be mature to guaranty that all necessary information is kept. When developing a
workflow, or when someone wants to re-use a previous execution, or explore new
orchestration directions, it becomes a limitation. Another consequence of such a
data-flow, is that past tasks/jobs cannot easily be queried and/or be used by
the global orchestrator to construct and alter workflow and orchestration
decisions.

In computational \textit{Materials Science}, simulation tasks are complex, often
requiring significant effort to integrate with general-purpose tools.
Consequently, several research groups have combined workflow executors with
molecular-scale simulators. With no exception, all delivered Python
interfaces/descriptions. For example, Custodian\cite{noauthor_custodian_nodate},
Atomate2\cite{MATHEW2017140}, AiiDA\cite{PIZZI2016218}, and
pyiron\cite{pyiron-paper} support high-throughput atomic/DFT calculations on
mainstream simulators within HPC environments. These platforms often allow MLIP
optimization and interaction with workflow orchestration.
FLAME\cite{noauthor_flame} and the more recent Autoplex\cite{noauthor_autoplex}
provide helpers to automate MLIP fitting for specific objectives. Some groups
have integrated active learning, including measures of MLIP prediction
reliability, such as FLARE\cite{noauthor_flare}, which uses sparse Gaussian
process regression for Bayesian uncertainty in workflow construction. 

\begin{table}[ht]
\label{label-features-workflow-managers}
  \centering
\tiny
\begin{tabularx}{\textwidth}{|l|l|X|X|X|c|c|c|}
\hline
Workflow & Executor & Metadata Storage & Data Storage & Supported Codes & DAG & Active Learning & pymatgen \\
\hline
Custodian & In-job (filesystem) & None (log/file inspection) & Filesystem & VASP, Gaussian, QE & $\times$ & $\times$ & \checkmark \\
\hline
Atomate2 & FireWorks / jobflow-remote & MongoDB & Filesystem / object store & VASP, QE, LAMMPS & \checkmark & $\times$ & \checkmark \\
\hline
AiiDA & Daemon (Slurm, PBS, LSF) & PostgreSQL (relational + graph) & File repository (object store) & VASP, QE, CP2K, LAMMPS, GPAW & \checkmark & $\times$ & \checkmark \\
\hline
pyiron & HPC schedulers (Slurm, PBS, LSF) & Project DB (SQLite optional) & HDF5 + filesystem & LAMMPS, VASP, GROMACS & \checkmark & $\times$ & $\times$ \\
\hline
FLAME & Python / Fortran routines & Filesystem (YAML configs) & Filesystem & NN potentials, MD, PES sampling & $\times$ & \checkmark & $\times$ \\
\hline
FLARE & Python & Files / lightweight DB & Filesystem & Sparse GP MLIP, MD & $\times$ & \checkmark & $\times$ \\
\hline
Autoplex & FireWorks / jobflow-remote & MongoDB & Filesystem / object store & ACE, EAM, MLIP benchmarking & \checkmark & $\times$ & \checkmark \\
\hline
AutoPot & Motoko (event-based) & Zeo (object-oriented DB) & Persistent object store (all intermediates) & MLIPs, LAMMPS + DFT refs & \checkmark & \checkmark & $\times$ \\
\hline
\end{tabularx}
\caption{Comparison of materials-science workflow frameworks with metadata/data storage, workflow preservation as a \textit{directed acyclic graph} (DAG), active learning, and pymatgen support. Disclaimer: this list was aggregated during early 2026, with large language models, in a field that keeps evolving rapidly.}
\end{table}

A summary of the possibilities offered by material sciences workflow managers
is proposed in table \ref{label-features-workflow-managers}. The type of storage,
the supported simulation codes, the retention of \textit{directed acyclic graph} (DAG)
for the workflow executions, active learning features and compatibility with the
very standard pymatgen library is indicated.

\begin{figure}[hbt]
    \centering
    \includegraphics[width=0.95\linewidth]{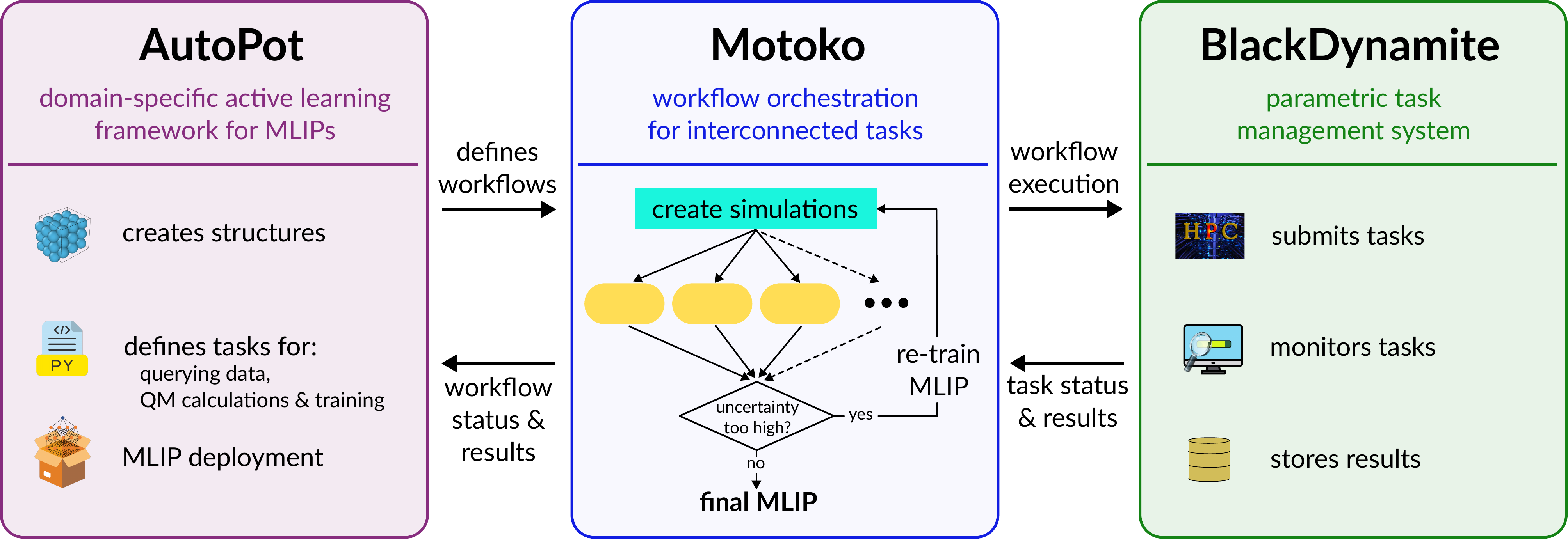}
    \caption{Schematic overview of the proposed framework for actively learning MLIPs during simulations}
    \label{fig:overview_autopot}
\end{figure}

In the current work, we will present a framework allowing MLIPs, to be obtained
through active learning, where tasks are 1) producing configurations, 2)
selecting configurations based on the MLIP's uncertainty, 3) single point
evaluation of energy, forces, and stresses, and 4) training the potential,
within complex iterative algorithms (see Figure \ref{fig:overview_autopot}). To this end, we propose a novel workflow
orchestrator, \textit{Motoko} \cite{anciaux_motoko} that keeps every information
produced by such tasks, orders metadata in an object database, while providing an
asynchronous python description of the global orchestration. \textit{AutoPot} is
then presented as a fully functional workflow allowing to obtain Moment Tensor
Potentials \cite{shapeev_moment_2016}, a class of MLIPs, by active learning
based on the D-optimality criterion \cite{settles_active_2009}. All details of
the orchestration are accessible within an open-source repository
\cite{hodapp_autopot}.
To our knowledge this is the only existing project allowing highly dynamic
orchestration focusing on D-optimality estimates. Also, the developments
presented below were conceived to allow rapid prototyping (flexibility for
testing new ideas) and an easy conversion to production mode, i.e. industrial
projects ran on HPC resources.

\section{Active learning for Moment Tensor Potentials\label{sec:active-learning}}

Before outlining the methodology, we establish the following notation. 
Consider a periodic configuration $\{\ur_i\} = \{\ur_i\}_{i=1}^N$ of $N$ atoms, representing the system at a given step during an atomistic simulation.
The neighborhood of the $i$-th atom is defined as the set $\scN = \{\ur_{ij}, z_i, \{ z_j \}\}$ of all relative atomic positions $\ur_{ij} = \ur_j - \ur_i$ between the atom located at $\ur_i$ and atoms $\ur_j$ located within a cut-off radius of a few lattice spacings from atom $i$, and the species of atom $i$, $z_i$, and its neighbors $\{ z_j \}$.
We assume that the total energy $\Pi$ of $\{\ur_i\}$ can be decomposed into per-atom energy contributions $\clE(\scN)$, so that
\begin{equation}\label{eq:total_energy}
    \Pi(\{\ur_i\}) = \sum_{\scN \in \{\ur_i\}} \clE(\scN).
\end{equation}

\subsection{Moment Tensor Potentials}

The per-atom energies $\clE$ of Moment Tensor Potentials (MTPs) \cite{shapeev_moment_2016,gubaev_machine_2018} are expressed as a linear combination of basis functions $B_\alpha$
\begin{equation}\label{eq:per-atom_energy}
    \clE(\scN) = \sum_\alpha \xi_\alpha B_\alpha(\scN),
\end{equation}
with learnable coefficients $\xi_\alpha$.
The basis functions are constructed from scalar contractions of the moment tensors
\begin{equation}\label{eq:mom-tens-desc}
    M_{\mu, \nu}(\scN)
    =
    \sum_{\ur_{i j} \in \scN} \sum_n c_{\mu n}(z_i, z_j) f_n\left(\left|\ur_{i j}\right|\right)
    \big( \underbrace{\ur_{ij} \otimes \cdots \otimes \ur_{ij}}_{\nu \text{ times}} \big),
\end{equation}
where the $c_{\mu n}$'s are free (nonlinear) parameters, and the $f_n$'s are Chebyshev radial basis functions smoothly vanishing at the potential cut-off.
In equation \eqref{eq:mom-tens-desc}, the indices $\mu$ and $\nu$ represent the moment tensor level given by $\operatorname{lev}M_{\mu,\nu} = 2 + 4\mu + \nu$.
An MTP with a level $\operatorname{lev}_{\rm MTP}$ is constructed from all basis functions obtained by all scalar contractions of the $M_{\mu, \nu}$'s satisfying 
$
    \operatorname{lev}_{\rm MTP} \geq \sum_{i=0}^{N_{\rm m}} \operatorname{lev}M_{\mu_i,\nu_i},
$
with $N_{\rm m}$ being the number of moment tensors involved in the contraction;
for example, a level-16 MTP contains 92 basis functions.

The parameters $\xi_\alpha$ and $c_{\mu n}$ are obtained from a minimization of the loss function
\begin{equation}\label{eq:loss}
    \scL =
    \sum_k \Bigg(
	w_{\rm e} \Big( \Pi_k - \Pi^{\rm dft}_k \Big)^2
	+
	w_{\rm f} \left( \sum_i \| \uf_{k,i} - \uf^{\rm dft}_{k,i} \|^2 \right) \\
    +
    w_{\rm s} \| \uusigma_k - \uusigma^{\rm dft}_k \|^2
    \Bigg)
    ,
\end{equation}
where $\Pi_k$, $\uf_{k,i}$, and $\uusigma_k$, are the MTP energies, forces, and stresses, of a configuration $k$ from the training set, $\Pi^{\rm dft}_k$, $\uf^{\rm dft}_{k,i}$, and $\uusigma^{\rm dft}_k$, are the corresponding quantities of the quantum-mechanical model on which the MTP is trained, and $w_{\rm e}$, $w_{\rm f}$, and $w_{\rm s}$, are the weights of the three different terms.

\subsection{Uncertainty quantification}

Suppose now that we are given an MTP that has been trained on some number of configurations.
We now want to check whether a \emph{new} configuration $\{ \ur_i \}^\ast$, not contained in the training set, should be added to the training set so that the MTP interpolates the quantum-mechanical energy at $\{ \ur_i \}^\ast$.

The algorithm that decides whether to add this configuration to our training set is called query strategy \cite{settles_active_2009}.
Here, we use the well-established D-optimality criterion for MTPs \cite{podryabinkin_active_2017}, as implemented in MLIP-2.
The D-optimality criterion is an error indicator and guarantees a stable potential, however, we remark that it is not necessarily a good quantification of the true error.
D-optimality estimates the MTP's uncertainty based on a scalar variable, the extrapolation grade $\gamma$ (it is called extrapolation grade because, mathematically, $\gamma$ has the meaning of a degree of extrapolation).

To compute $\gamma$, consider an MTP with $m$ parameters and an active set containing $m$ configurations.
We now define the $m \times m$ Jacobian matrix
\begin{equation}\label{eq:Jacobian}
    \uuA
    =
    \begin{pmatrix}
        \frac{\partial \Pi(\{ \ur_i \}_1; \mtheta)}{\partial \mtheta_1} & \cdots & \frac{\partial \Pi(\{ \ur_i \}_1; \mtheta)}{\partial \mtheta_m} \\
        \vdots & \ddots & \vdots \\
        \frac{\partial \Pi(\{ \ur_i \}_m; \mtheta)}{\partial \mtheta_1} & \cdots & \frac{\partial \Pi(\{ \ur_i \}_m; \mtheta)}{\partial \mtheta_m}
    \end{pmatrix}
    ,
\end{equation}
where $\mtheta = \{ \xi_\alpha$, $c_{\mu n} \}$ includes all linear and nonlinear parameters.
The extrapolation grade $\gamma$ is then defined as the maximum change in the determinant of $\uuA$ when replacing any of the configurations from the active set with $\{ \ur_i \}^\ast$.
Computing all determinants individually is unnecessary.
Instead, $\gamma$ can be efficiently calculated as
\begin{equation}\label{eq:grade}
    \gamma
    =
    \underset{i}{\operatorname{max}} \vert c_i \vert,
    \qquad
    \text{with}
    \;
    \uc
    =
    \begin{pmatrix}
        \frac{\partial \Pi(\{ \ur_i \}^\ast; \umtheta)}{\partial \mtheta_1} & \cdots & \frac{\partial \Pi(\{ \ur_i \}^\ast; \umtheta))}{\partial \mtheta_m}
    \end{pmatrix}^\sT
    \uuA^{-1}
    .
\end{equation}
We remark that the training set typically contains many more configurations than parameters, making $\uuA$ overdetermined.
Therefore, the $m$ neighborhoods are selected so that linear independence of the column vectors of $\uuA$ is maximized;
this is done using the maxvol algorithm of \citet{olshevsky_how_2010}.

To decide whether $\{ \ur_i \}^\ast$ should be added to the training set, \citet{novikov_mlip_2021} defined the following classification of extrapolation grades:
\[
    \begin{aligned}
        &\gamma \le 1 && \text{indicates interpolation,} \\
        1 < &\gamma \le 2 && \text{indicates accurate extrapolation,} \\
        2 < &\gamma \le 10 && \text{indicates still reliable extrapolation,} \\
        10 < &\gamma && \text{indicates risky extrapolation}.
    \end{aligned}
\]

\subsection{Learning protocols}
\label{sec:learning_protocols}

Having the scalar uncertainty $\gamma$ well-defined, there are now multiple ways to implement active learning algorithms in order to assemble a good training set.
Below, we will describe two of them that have proven particularly useful for constructing unbiased training sets for random alloys \cite{hodapp_machine-learning_2021,novikov_ai-accelerated_2022,moitzi_ab_2024}, and are implemented in the initial version of AutoPot.

The first algorithm addresses the problem of having a large set of candidate configurations $\scT_{\rm cand}$ that we cannot all afford to compute with a quantum-mechanical model.
To that end, Algorithm \ref{algo:selection} iterates through all the configurations in the candidate set $\scT_{\rm cand}$, and checks whether the extrapolation grade exceeds a threshold indicating that the corresponding configuration should be added to the training set.

\begin{algorithm}[hbt]
    \SetAlgoSkip{bigskip}
    \LinesNumbered
    \SetKwInput{Input}{Input}
    \SetKwInput{Output}{Output}
    \caption{Selection}
    \label{algo:selection}
    \Input{MTP, training candidate set $\scT_{\rm cand}$, $\gamma_{\rm threshold}$}
    $\scT_{\rm selected} \; \leftarrow \; \{\}$; \tcp{initialize set of selected configurations to be added to the training set}
    \For{$\{ \ur_i \}^\ast \in \scT_{\rm cand}$}{
        Compute $\gamma$ of $\{ \ur_i \}^\ast$ \eqref{eq:grade}; \\
        \If{$\gamma > \gamma_{\rm threshold}$}{
            Add $\{ \ur_i \}^\ast$ to $\scT_{\rm selected}$; \\
            Update the matrix $\uuA$ \eqref{eq:Jacobian}; \\
        }
    }
    Run a single-point calculation on each configuration in $\scT_{\rm selected}$, and add them to the training set; \\
    Re-train the MTP on the updated training set; \\
    \Output{New MTP}
\end{algorithm}

When making selections from large sets of training candidates, Algorithm \ref{algo:selection} is typically called iteratively, with decreasing $\gamma_{\rm threshold}$'s as input.
The reasoning behind this strategy is as follows:
suppose that training candidates are ordered with increasing extrapolation grade, i.e. increasing distance from the training set.
Then, after adding the first candidate to $\scT_{\rm select}$ and updating the matrix $\uuA$, the next candidate may still have an extrapolation above $\gamma_{\rm threshold}$ and will be added to $\scT_{\rm select}$.
In practice, however, it suffices to add the candidate with the \emph{highest} extrapolation to $\scT_{\rm select}$ so that all other candidates are interpolated by the potential.
This strategy therefore ensures that the chance of selecting too many similar configurations is strongly reduced, keeping the training set small without sacrificing accuracy (cf. \cite{hodapp_machine-learning_2021}).

The second algorithm addresses the problem of a training candidate set not being representative enough to ensure a certain level of accuracy of a simulation using the MTP, or even to ensure the stability of the simulation.
In such cases, we need to sample the training configurations on-the-fly while running the simulations, e.g., structural relaxation, molecular dynamics, Monte Carlo simulations, etc.
Algorithm \ref{algo:on_the_fly_selection} exemplifies the main steps of such a sampling procedure.

\begin{algorithm}[hbt]
    \SetAlgoSkip{bigskip}
    \LinesNumbered
    \SetKwInput{Input}{Input}
    \SetKwInput{Output}{Output}
    \caption{On-the-fly selection}
    \label{algo:on_the_fly_selection}
    \Input{MTP, configurations $\{ \ur_i \}_j$ on which to run the simulations, thresholds $\gamma_{\rm min}$ and $\gamma_{\rm max}$}
    $\scT_{\rm cand} \; \leftarrow \; \{\}$; \tcp{initialize set of training candidates}
    \While{uncertain configurations are found during the simulations}{
        \For{$\{ \ur_i \}^\ast \in \{ \ur_i \}_j$}{
            \While{simulation is not converged}{
                Evolve $\{ \ur_i \}^\ast$; \\
                Compute the extrapolation grade $\gamma^\ast$ of $\{ \ur_i \}^\ast$; \\
                \uIf{$\gamma^\ast > \gamma_{\rm min}$}{
                    Add $\{ \ur_i \}^\ast$ to $\scT_{\rm cand}$; \\
                }
                \ElseIf{$\gamma^\ast > \gamma_{\rm max}$}{
                    Abort the simulation; \\
                }
            }
        }
    }
    Add the most representative configurations from $\scT_{\rm cand}$ to the training set and construct a new MTP using Algorithm \ref{algo:selection} with $\gamma_{\rm threshold} = \gamma_{\rm min}$; \\
    \Output{New MTP}
\end{algorithm}
  
One challenge when implementing Algorithm \ref{algo:selection} and \ref{algo:on_the_fly_selection} is the communication between the MLIP, ab initio, and simulation codes, that needs to happen in a concurrent manner because their execution may follow a highly heterogeneous pattern.
For example, in Algorithm \ref{algo:on_the_fly_selection}, we may sample from many (possibly thousands) of simulations each triggering new single-point calculations at different points in time.
In such cases, we do not want to wait until \emph{all} simulations are finished.
Instead, we want to run the single-point calculations immediately, while other simulations continue running.
This requires a workflow manager that allows to place conditions on individual simulations based on which another simulation is started.

\begin{figure}[ht]
  \centering
  \includegraphics[width=.5\linewidth]{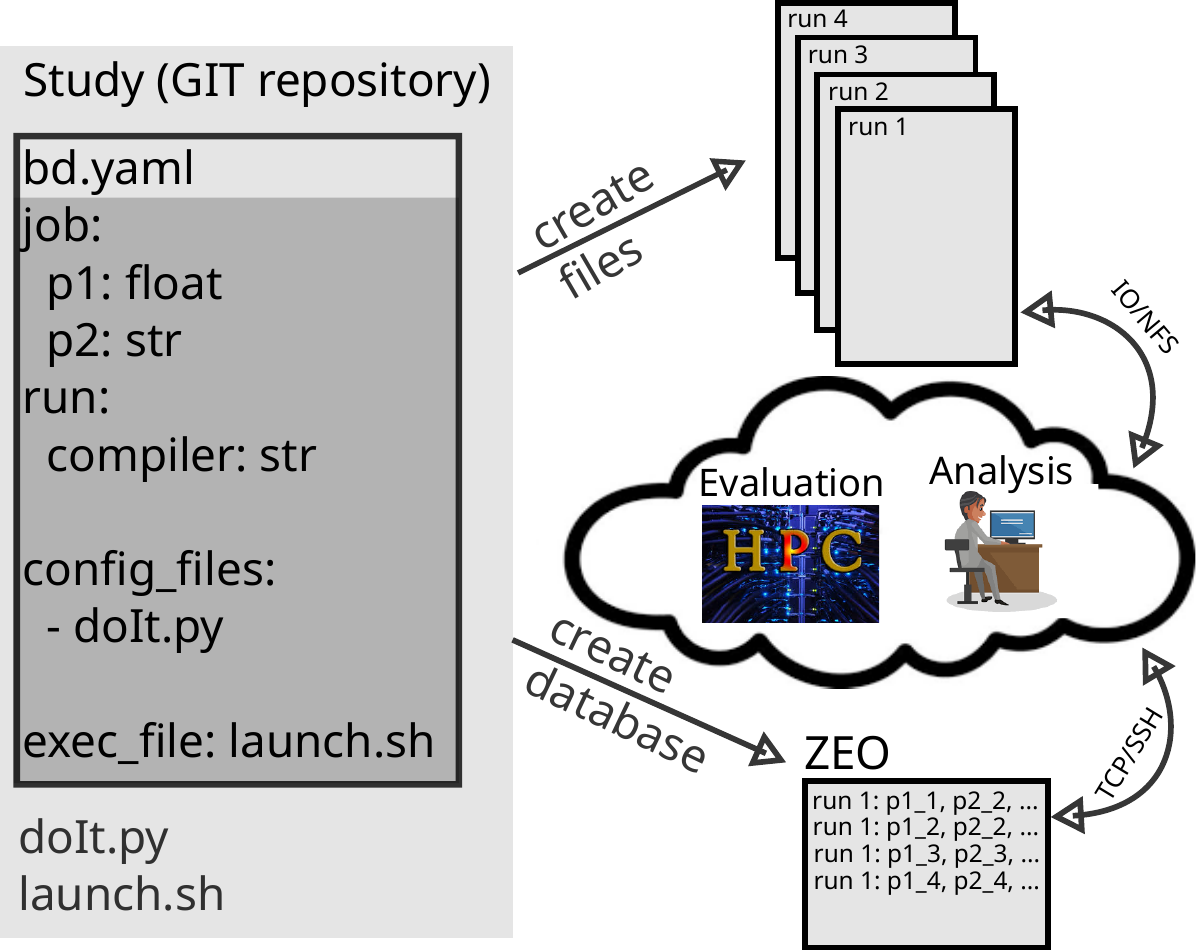}
  \caption{\label{fig:blackdynamite}Schematic of the services offered by the Blackdynamite parametric study framework. A study/task constists in a `yaml` file describing the parametric space, and the attached script files necessary to launch a simulation. This information, light, and saved in a GIT repository, is usually enough to re-run a parametric study . For execution, Blackdynamite takes care of the (remote) reservation of storage\&computing  resources and finally stores all outcome into an object database. Remote queries to the database later allow analysis of the results, and remote file access to any produced file.}
\end{figure}

The main aim of the present work is to demonstrate that the above challenges can be addressed using a novel workflow orchestrator based on the BlackDynamite parametric study manager.
As shown in Figure \ref{fig:blackdynamite}, \textit{Blackdynamite} describes a task (aka. a parametric study) using a `yaml` file with attached script files that enable the execution
of simulations by varying the parameters within a parametric space. \textit{BlackDynamite} organizes access to storage and computational resources (remote/HPC), as will be described in section \ref{section:tasks}. In what follows we will describe how tasks from various
simulations are orchestrated allowing energy calculations to trigger training tasks,
with a natural description of complex and dynamical workflows.

\section{\textit{Motoko} workflow manager}

The workflow described in this work is managed with the new versatile workflow
manager \textit{Motoko}, which is part of a collection of software, designed to
facilitate exploration of defined parametric spaces by executing tasks and
keeping records for reproducibility. \textit{Motoko} component enables event-based as
well as asynchronous-type orchestration to be written in Python. Concretely, one
can execute asynchronous routines following a dependency graph, either described
as asyncio Python routines, or as a collection of events triggering actions. The
actions can naturally be simple routines. But they can also be pre-defined
complex tasks. With every task execution, all the information relative to input
parameters, software context, as well as output are preserved in a database,
enabling structured queries. Such tasks can be run locally,
on distant machines, and/or on high performance computing clusters by submitting
tasks to job schedulers (e.g. slurm). More details will follow in next section.

Contrary to other workflow managers \cite{flyte, kubeflow_pipelines,
mlflow,airflow, zenml, ,metaflow, scikit_learn}, the storage is not ephemeral,
i.e. every dataset produced by a task execution is kept locally on the
machine that produced it (e.g. the HPC computing cluster), and/or on an
object-database with remote access possibilities.
It allows making queries to select tasks based on their input
and/or output, during the workflow execution, but also later during results
processing, with fullest possible details, aiming at enforcing reproducibility.
It also allows to explore new training routes, without having to re-run the
entire workflow, making exploration of new ideas flexible.

In Motoko, collections of tasks can be generated, queried, and manipulated
easily. One example of an action-event relying on a collection of tasks would be
described by: ``\textit{When all single-point calculations are finished,
proceed to the MLIP training phase}''. In the above, the event condition
will be met when no more single-point calculation tasks keep running, while the action
will create a new \textit{training} task.

When the entire workflow execution is made on a single computing resource (e.g. one
laptop or one HPC cluster), all the execution and database files should be
contained in the workflow directory. Making a compressed archive (tar, zip)
is sufficient to pack all relevant information, enabling in depth
reproducibility of executing the workflow, as well as the produced results.
If one wants to export the final training set to standard repositories
like \textit{ColabFit} \cite{vita_colabfit_2023}, a query to the database
is sufficient. For instance, it would also be possible to export only a
subset of the configurations based on the associated extrapolation grade.

\subsection{Tasks\label{section:tasks}}

A task can be viewed as a parametric study, where the output varies according to
input parameters. One key feature of the \textit{BlackDynamite} (BD) framework,
is the description and manipulation of parametric studies and
spaces \cite{blackdynamite}. Therefore, each task needs to be characterized to
control and monitor execution from chosen parameters. A BD task
is provided a parametric space which is described by a hierarchical \textit{yaml} file, and based on
that information, BD organizes the execution of tasks and produced files while keeping record of
input parameters. Input parameters and chosen results are collected and organized in databases
to enable complex queries. The ZODB/ZEO \cite{zodb, zeo} object-database
serve as the storage for data and metadata, and 
accesses/queries to these are ensured by \textit{BlackDynamite} API routines or
with command line clients.
  The execution of a collection of tasks can be local (default) and therefore sequential,
  but can also be submitted to any HPC cluster job schedulers (e.g. Slurm) allowing 
  remote executions transparently.  
Furthermore, BD keeps track of all files and directories so that it is almost
impossible to lose produced data. BD also provides remote access to
files and database content, particularly adapted to the usage of multiple HPC
computing resources. After completion of the tasks, the analysis is facilitated
through automatic retrieval/conversion of data with \textit{Numpy} arrays.

\subsection{Orchestrator}

A \textit{Motoko} orchestrator is described with an asynchronous python program, to be placed in
a directory containing a \textbf{motoko.yaml} file, which essentially declares
the tasks, and the orchestrator main function:

\begin{minted}[fontsize=\scriptsize, bgcolor=gray!10]{yaml}
task_managers:
  task1:

  task2:

orchestrator: orchestrator.main
\end{minted}

Here, \texttt{task1} and \texttt{task2}, are tasks, i.e. an external script
(usually a set of python functions) that serves as a template to create actual
executions. The orchestrator can start/stop new executions of any declared task.
The orchestrator role, is to decide when to execute new tasks, in reaction to
results of others. For example, by launching specific instances of
\texttt{task2} by choosing its parameters based on the output of \texttt{task1}
instances.

The asynchronous orchestrator is usually placed in a python script called
\textbf{orchestrator.py}, while \textbf{main} is the function starting the
workflow. The folder structure of a Motoko workflow then looks as follows:

\begin{minted}[fontsize=\scriptsize, bgcolor=gray!10]{bash}
workflow/
    orchestrator.py
    motoko.yaml
    task1/
        run.sh
        doIt.py
        bd.yaml
    task2/
        run.sh
        doIt.py
        bd.yaml
\end{minted}

Here, the \texttt{run.sh} files are bash scripts calling the Python functions of the corresponding task provided in the \texttt{doIt.py} files.
The asynchronous execution can be demanded with the command \textbf{motoko orchestrator start} from the \texttt{workflow/} folder.

In the following sections, details are provided on how to declare the entire
Motoko workflow, from tasks up to the asynchronous routines realizing the
orchestration for our MLIP training.

\section{Autopot workflow}

For MLIPs conceived with active machine-learning, it is neither necessary nor
desired to explore the entire parametric space (we cannot compute all the
energies for an infinite number of configurations). Instead, adequate selections
of training configurations should be performed dynamically during the global
workflow execution. The goal is to find a small, affordable set of training
configurations to achieve the desired accuracy. \textit{AutoPot} is a workflow
orchestrator, following Motoko's framework, that realizes active learning as
described in Section \ref{sec:active-learning}.

\begin{figure}[hbt]
    \centering
    \includegraphics[width=0.8\textwidth]{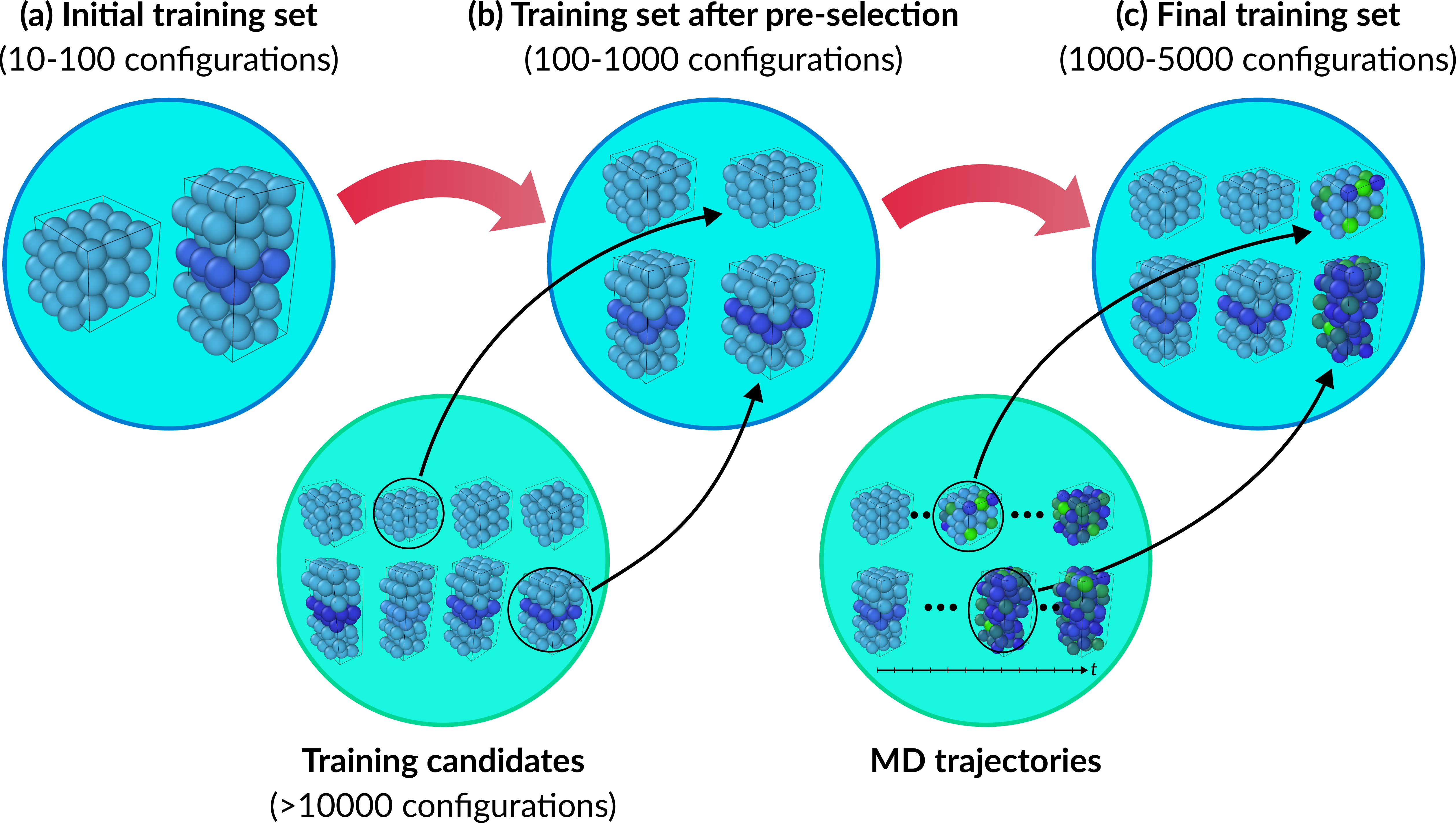}
    \caption{Schematic of the active learning workflow implemented in AutoPot}
    \label{fig:workflow_schematic}
\end{figure}

The AutoPot workflow combines the Algorithms from Section \ref{sec:learning_protocols}, and is schematically depicted in Figure \ref{fig:workflow_schematic}.
In this algorithm, the training set is constructed in three steps:
\begin{itemize}
    \item[(a)]
    Starting from an initial training set,
    \item[(b)]
    AutoPot adds the most uncertain configurations from a large set of training candidates using Algorithm \ref{algo:selection},
    \item[(c)]
    followed by a selection of training configurations from MD trajectories using Algorithm \ref{algo:on_the_fly_selection}.
\end{itemize}

As a rule of thumb, configurations composing the initial training set and the training candidate set should contain atomic neighborhoods relevant to the planned application.
For example, for our tests in Section \ref{sec:production_mode} related to dislocation plasticity, we use two types of configurations: bulk configurations, and configurations with a stacking fault (cf., Figure \ref{fig:workflow_schematic}).
The initial training set and the candidate set are then constructed by randomly perturbing atoms from their ideal lattice sites.

Another possibility is to use an existing dataset from some database.

In AutoPot, the above workflow is cast into a Motoko workflow.
This requires defining BlackDynamite tasks and connecting them in the Motoko orchestrator.
This will be described in the remainder of this section.

\subsection{Tasks}

To realize the above workflow, \textit{Autopot} connects four tasks (\textbf{sp\_calc}, and \textbf{train}, \textbf{select}, and \textbf{md\_select}).
In more details, the first two tasks handle the DFT calculations and the training of the potential:
\begin{enumerate}
  \item \textbf{sp\_calc}: \textit{Single point calculation} evaluating the energy, forces, and stresses, for a given atomic configuration.
        Two versions of \texttt{sp\_calc} have been implemented, one that uses the \texttt{EAMCalculator} from ASE as a reference for testing.
        The other version runs a VASP calculation.
  \item \textbf{train}: \textit{Training the MLIP} that minimizes the loss function \eqref{eq:loss} for a given training set.
        Such a task executes the command \texttt{mlp train} from MLIP-2.
\end{enumerate}
Selecting uncertain configurations from the training candidate set and the MD simulations is handled by the remaining two tasks:
\begin{enumerate}[resume]
  \item \textbf{select}: \textit{Selection of configurations} by evaluating the extrapolation grade. It follows Algorithm \ref{algo:selection} which
        will execute the \texttt{select-add} command from MLIP-2 to populate a file with configurations on which to perform a single-point calculation.
  \item \textbf{md\_select}: \textit{Molecular dynamics simulations} used to select configurations. It runs Algorithm \ref{algo:on_the_fly_selection} with
        Langevin dynamics, as implemented in ASE \cite{hjorth_larsen_atomic_2017}.
        After each MD step, it executes \texttt{calc-grade} from MLIP-2 to compute the extrapolation grade for the current configuration, and populates a file with those configurations exceeding the extrapolation threshold on which to perform a single-point calculation.
\end{enumerate}

These four tasks are described using BlackDynamite's framework.
Task execution is handled by a Python script, e.g., for \textbf{sp\_calc} running a VASP calculation with ASE it would call a function like:

\begin{minted}[bgcolor=gray!10]{python}
def doIt(run, job):
    incar_fname = job.incar_fname
    cfg_fname = job.cfg_fname

    cfg = read(cfg_fname)
        
    calc = create_vasp_calc(incar_fname)
    cfg.calc = calc

    # Compute energy, forces, and stress
    energy = cfg.get_potential_energy()
    forces = cfg.get_forces()
    stress = cfg.get_stress()

    # Push quantities to blob
    run.pushQuantity(energy, 0, "energy")
    run.pushQuantity(forces.flatten(), 0, "forces")
    run.pushQuantity(stress, 0, "stress")
\end{minted}

The parameter space associated which each of these tasks is given in a
\texttt{.yaml} file, as shown in Figure \ref{fig:yaml_files}. 
The parametric space is two-fold: 1) the job space describes the physical 
quantities or model properties controlling the task. 2) the run space provides
the opportunity to describe parameters regarding the actual execution.
For instance,  the temperature is a job parameter while the slurm options or the MD software are usually run parameters.
In summary,
selections (\textbf{select} or \textbf{md\_select}) identify the uncertain
configurations, and single-point calculations (\textbf{sp\_calc}) compute the
energies, forces, and stresses, of selected configurations in order to update the
training set. Each time the training set is determined, a \texttt{train} task
will produce a new MLIP by minimizing the loss function. This sequence of task
executions is performed until no new uncertain configuration is found/created from 
the selection tasks.

\begin{figure}[ht]
\hrule
\begin{minipage}{0.24\textwidth}
\begin{minted}[fontsize=\scriptsize, bgcolor=gray!10]{yaml}
task: sp_calc

job:
  cfg_fname: str
  incar_fname: str

run:
  slurm_options: list
\end{minted}
\end{minipage}
\hfill
\begin{minipage}{0.24\textwidth}
\begin{minted}[fontsize=\scriptsize, bgcolor=gray!10]{yaml}  
task: train

job:
  pot_fname: str
  ts_fname: str
  sp_calc_ids: list
  atom_types: str
  train_args: dict

run:
  slurm_options: list
\end{minted}
\end{minipage}
\hfill
\begin{minipage}{0.24\textwidth}
\begin{minted}[fontsize=\scriptsize, bgcolor=gray!10]{yaml}
task: select

job:
  grade: float
  training_candidate_sets: list
  pot_fname: str
  ts_fname: str
  atom_types: str

run:
  slurm_options: list
\end{minted}
\end{minipage}
\hfill
\begin{minipage}{0.24\textwidth}
\begin{minted}[fontsize=\scriptsize, bgcolor=gray!10]{yaml}
task: md_select

job:
  cfg_fname: str
  pot_fname: str
  ts_fname: str
  atom_types: str
  md_args: dict
  select_args: dict

run:
  md_software: str
  slurm_options: list
\end{minted}
\end{minipage}
\hrule

\caption{\texttt{.yaml} files of the tasks implemented in AutoPot.
For the \texttt{sp\_calc} version that runs a VASP calculation, \texttt{pot\_fname} corresponds to the INCAR file.
For the \texttt{train} tasks, the id's of \texttt{sp\_calc} tasks can be provided as a list in order to add the calculated energies, forces, and stresses, from those \texttt{sp\_calc} tasks to the training set.}
\label{fig:yaml_files}
\end{figure}

\subsection{Orchestration}

In order to complete the workflow description, the orchestration is given
as an asynchronous python program, to be placed in a directory
containing a \textbf{motoko.yaml} file. This file registers the tasks, as well as
the orchestrator's main function:

\begin{minted}[fontsize=\scriptsize, bgcolor=gray!10]{yaml}
task_managers:
  select:

  md_select:

  sp_calc:

  train:

orchestrator: orchestrator.main
\end{minted}

Essentially, the workflow is described with python functions that are declared
as \textbf{async}, making them not necessarily executed immediately. The obvious
reason being that the orchestrator may trigger the creation of tasks, the execution scheduling,
and finally the storage of results which will take time. Before the workflow
can get a chance to resume execution, the \textit{async} declaration offers it the possibility to treat other 
orchestration duties asynchronously. An example of such a routine would be:

\begin{minted}[bgcolor=gray!10]{python}
@event
async def spawn_sp_calc_task(workflow, cfg_fnames):
    created_calc_runs = await workflow.sp_calc.createTask(cfg_fname=cfg_fnames)
    return created_calc_runs
\end{minted}

In the example above, configuration filenames are used to create one \textbf{sp\_calc} task per configuration with the \textbf{createTask} call, i.e., it would call the \textbf{doIt} function exemplified in the previous section for each element in the list \textbf{cfg\_names}.
The \textbf{await} keyword specifies that this is an asynchronous call which needs
another execution to finish before resuming execution of the orchestration.
After completion, the return variable \textbf{created\_calc\_runs} is a
collection of executed (BD) tasks that may be queried for properties, state,
results, or even remotely produced files.

\begin{figure}[t!]
  \begin{center}
    \includegraphics[width=0.97\textwidth]{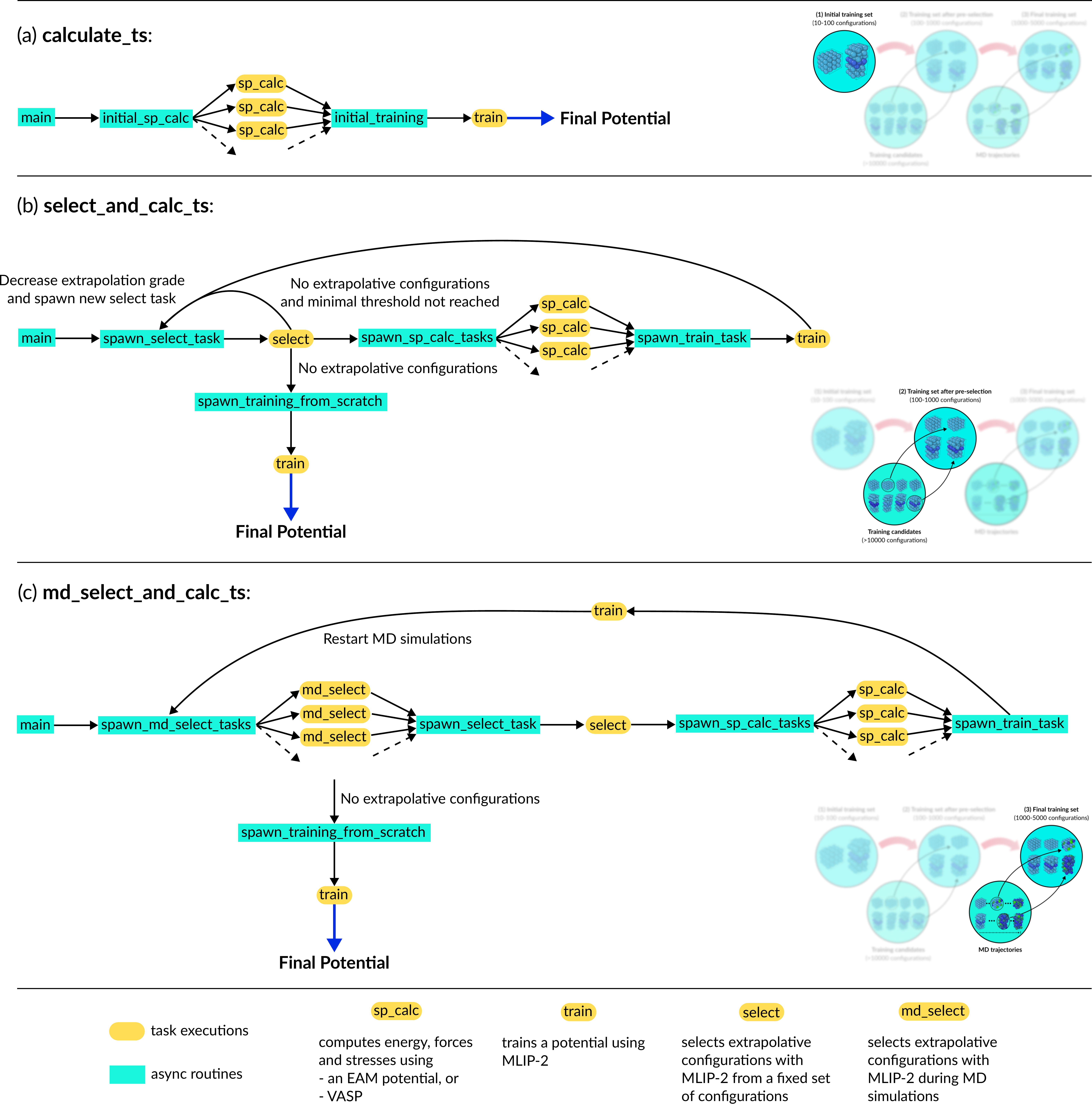}
\end{center}
    \caption{Workflow stages implemented in AutoPot.
    The cyan boxes are \textit{async} routines given in the file \textbf{orchestrator.py}, while yellow boxes refer  to \textit{task} executions}
    \label{fig:autopot_workflows}
\end{figure}

We have decomposed the full schematic AutoPot workflow from Figure \ref{fig:workflow_schematic} into the three stages shown
in Figure \ref{fig:autopot_workflows}, each of them taking an initial training set as input and return a trained potential:
\begin{itemize}
    \item[(a)]
    \textbf{calculate\_ts}
    produces a training set from an initial set of configurations.
    \item[(b)]
    \textbf{select\_and\_calc\_ts}
    will sample the training set from a training candidate set based on a varying extrapolation threshold (Algorithm \ref{algo:selection}).
    \item[(c)]
    \textbf{md\_select\_and\_calc\_ts}
    will sample the training set from MD simulations on a chosen set  of configurations (Algorithm \ref{algo:on_the_fly_selection}).
\end{itemize}
Once uncertain configurations are found in any of the selection stages, \textbf{select} and \textbf{md\_select}, corresponding \textbf{sp\_calc} runs will be spawned.
Again, when all \textbf{sp\_calc} runs are finished, a new \texttt{train} run will add all new configurations to the training set and retrain the potential.

The \textbf{select} stage could produce an unnecessarily large number of training configurations for very large training candidate sets when selecting all configurations above a small extrapolation grade between 1 and 2 at once (cf. Section \ref{sec:learning_protocols}).
To avoid this, we iteratively spawn \textbf{select} tasks, first with a very large extrapolation threshold (e.g. >\,100\,000), then with some smaller extrapolation threshold (e.g. >\,10\,000), etc., and finally with an extrapolation threshold between 1 and 2 (corresponding to accurate extrapolation).
This ensures that configurations selected for training are sufficiently distinct from each other.

The three stages can be called standalone or in sequence from the orchestrator.
The behavior of the full orchestrator, as seen in Figure \ref{fig:full_workflow},
can be controlled by the following arguments:

\begin{minted}[fontsize=\scriptsize, bgcolor=gray!10]{bash}
  --potential 08.mtp
                        interatomic potential to start with
  --fresh_potential 08.mtp
                        untrained interatomic potential to be used for training from scratch after collecting all training configurations
  --training_set ts.cfg
                        initial training set
  --calculate_ts        calculate the initial training set and train the potential on it
  --atom_types W=0,Ta=1,..., -a W=0,Ta=1,...
                        mapping of atomic types to numbers according to convention used in the potential code
  --select              pre-select configurations from the candidate set
  --grades [10000 1000 100 ...], -g [10000 1000 100 ...]
                        Maximum extrapolation grades above which iteratively configurations are selected for training
  --training_candidate_set tcs.cfg
                        Set with the training candidates
  --md_select           select configurations from MD simulations
  --md_configurations configurations.xyz
                        file with atomic configurations on which to run the MD simulations. 
                        If not given, run MD on each configuration from the training set
\end{minted}

\noindent For instance, by running the workflow with the command line:
\begin{minted}[fontsize=\scriptsize, bgcolor=gray!10]{bash}
motoko orchestrator start --potential 08.mtp --training_set init_training_cfgs.xyz --calculate_ts --atom_types Mo=0\
                          --md_select --md_configurations md_cfgs.xyz 
\end{minted}

We have built in some flexibility by allowing to either choose \texttt{----select} and \texttt{----md\_select} individually, or a combination of both. If both options are chosen, then the workflow always starts with the \textbf{select} stage.
When all MD simulations are finished without triggering new single-point calculations, the orchestrator will terminate.
After completion of the workflow, the MTP file and training set from the last \texttt{train} task can be accessed with the following
python script:
\begin{minted}[fontsize=\scriptsize, bgcolor=gray!10]{python}
wf = Workflow("motoko.yaml")

# get the last task execution run
last_run = wf.train.connect().runs[wf.train.connect().runs_counter - 1]

# get the running directory for that run
run_path = os.path.join(
    wf.train.study_dir, f"BD-train-runs/run-{last_run["id"]}"
)

# obtain a full path name for the potential and training set
mtp_fname = os.path.join(
    run_path, last_run.train_args["trained_pot_name"]
)
ts_fname = os.path.join(
    run_path, last_run.train_args["ts_name"]
)
\end{minted}

\section{AutoPot in production mode}
\label{sec:production_mode}

We now test AutoPot in production mode by constructing two MTPs, one for bcc W, and one for Mo-Ta random alloys.
We benchmark the trained potentials on material properties relevant for nanoscale plasticity, such as elastic constants, stacking fault energies, and several key properties of dislocations.
There is plenty of DFT data available for this combination of materials and properties which makes it ideal for benchmarking.

We first describe the details of the training protocol that are the same for both potentials.
As an ab initio model we use DFT with a setup close to the one used in \cite{hodapp_operando_2020}:
we employ the Vienna Ab initio Simulation Package (VASP) \cite{kresse_efficient_1996} with plane-wave basis sets, the projector-augmented wave (PAW) pseudopotential method \cite{blochl_projector_1994,kresse_ultrasoft_1999}, and the PBE method \cite{perdew_generalized_1996} to calculate the exchange-correlation functional;
we further use an energy cutoff of 400\,eV, a Gaussian smearing of 0.08\,eV, and a minimum $\bmk$-point spacing of 0.15\,\AA$^{-1}$.
Electronic relaxation is performed using the preconditioned minimal residual method.
A configuration is considered as converged when the energy difference between two iterations is less than 5\,$\times$\,10$^{-7}$\,eV.

As an interatomic potential, we use an MTP of level 16 with the cut-off radius set to 5\,\AA.
Unless specified otherwise, we run AutoPot using the default options given in the previous section.
The examples presented in this section have been run on the Vienna Scientific Cluster (VSC-5) provided by the Austrian Scientific Computing (ASC) facilities.

\subsection{MTP for W}

We choose to train our potential on two types of configurations, bulk configurations with 54 atoms per supercell, and configurations containing a 1/4\,[111] stacking fault with 72 atoms per supercell.
For each type of configuration, we first create a very large set of training candidates with 10\,000 samples in their ground state.
A Cauchy strain is applied to each sample, extracted from a uniform distribution with the minimum and maximum strain components set to $\pm$\,2\;\%.
Moreover, the atomic positions of each strained sample are randomly perturbed using a Gaussian distribution with a standard deviation of 0.02\,\AA.

We start from a training set containing 20 configurations (10 bulk and 10 stacking fault configurations).
We then perform a pre-selection of training configurations from the training candidate set using the \texttt{select} task.
That is, we iteratively add configurations with high extrapolation grades by gradually lowering the threshold (cf. Algorithm \ref{algo:selection}) until the extrapolation grades of all configurations in the training candidate set are $<2$.
Afterwards, we sample the remaining training configurations from two NVT MD trajectories at 100\,K starting from one bulk and one stacking fault configuration in the ground state.

\begin{figure}[hbt]
    \centering
    \includegraphics[width=0.65\textwidth]{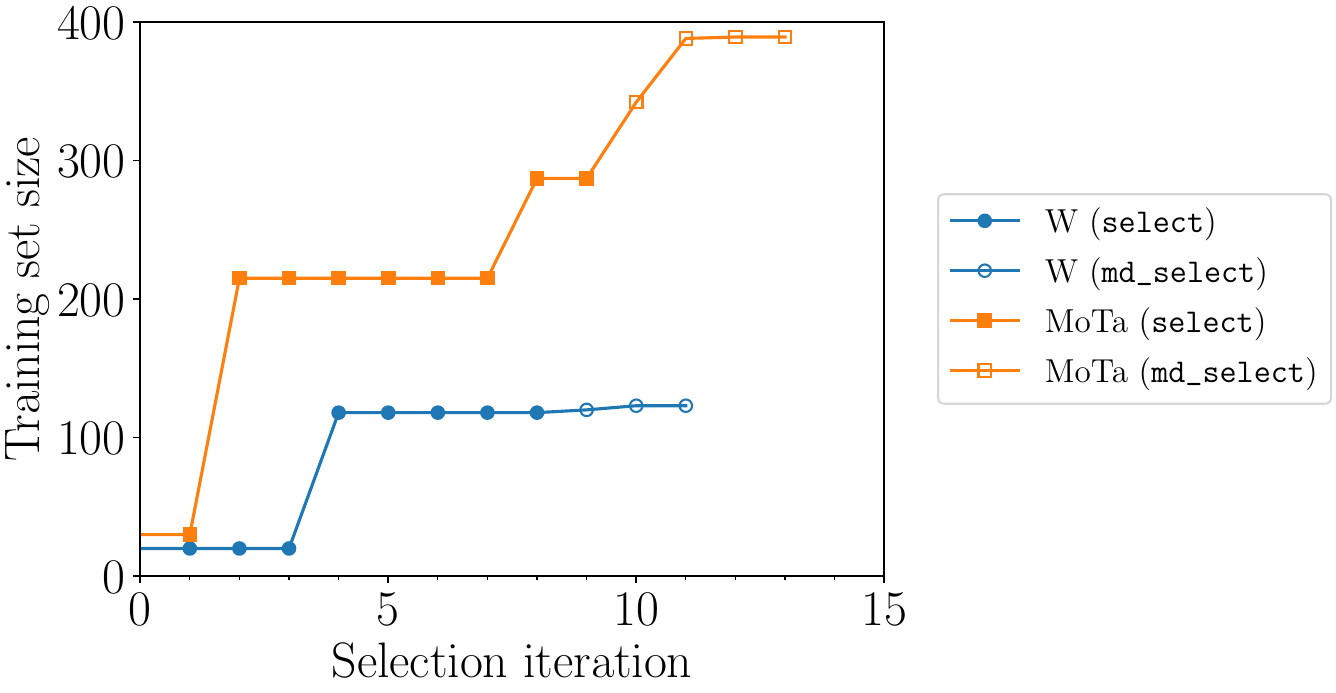}
    \caption{Evolution of training set size (number of configurations contained in the training set) of the MTPs for W and Mo-Ta as a function of the selection iteration}
    \label{fig:selected_cfgs}
\end{figure}

In Figure \ref{fig:selected_cfgs}, we show the evolution of the training set size as a function of the selection iteration.
The total number of configurations selected by AutoPot is 123.
Almost all of the configurations (118) are added during the third iteration of the pre-selection phase (when $\gamma_{\rm threshold} = 10\,000$).
From the MD simulations, only a small number of 5 configurations is selected for training.
This is somewhat expected and demonstrates the correct functioning of the AutoPot workflow:
configurations appearing in MD simulations at 100\,K are not too far from the ground state and are already sufficiently well covered by the large training candidate set of 10\,000 samples per configuration type.

The MTP training errors are shown in Table \ref{tab:training_errors}.
The errors are well within the range known to be sufficient for MTPs to predict elastic and defect properties in unary metals with near-DFT accuracy, i.e., of the order of 1\,meV for per-atom energies, of the order of 10\,meV/{\AA} for forces, and of the order of 10$^{-1}$\,GPa for stresses (cf. e.g,\cite{hodapp_operando_2020,mismetti_automated_2024}).

\begin{table}[hbt]
    \centering
    \begin{tabular}{|c|c|c|c|c|c|c|}
        \hline
        \multirow{2}{*}{MTP} & Energy MAE & Energy RMSE & Force MAE & Force RMSE & Stress MAE & Stress RMSE \\
                             & [meV/atom] & [meV/atom]  & [meV/\AA] & [meV/\AA]  & [GPa]      & [GPa] \\ \hline\hline
        W    & 0.50 & 0.78 & 33.5 & 64.0 & 0.064 & 0.216 \\ \hline
        MoTa & 2.18 & 2.91 & 39.0 & 76.2 & 0.090 & 0.291 \\ \hline
    \end{tabular}
    \caption{Mean absolute error (MAE) and root mean square error (RMSE) of the MTPs for W and MoTa}
    \label{tab:training_errors}
\end{table}

The MTP is then benchmarked on various properties relevant for studying dislocation plasticity, namely, the lattice constant, elastic constants, and the 1/4\,[111] unstable stacking fault energy;
the simulation details for all considered properties are given in Appendix \ref{sec:sim_details}.
Overall, the agreement with DFT is very good, as shown in Table \ref{tab:MTP_errors}.
The MTP error on all quantities is less than 10\,\%, comparable to other highly accurate MLIPs for W (e.g., \cite{byggmastar_machine-learning_2019,hodapp_operando_2020}).

\begin{table}[hbt]
    \centering
    \begin{tabular}{|c|c|c|c|c|c|}
        \hline
        \multirow{2}{*}{Method} & $a_0$ & $C_{11}$ & $C_{12}$ & $C_{44}$ & $\Pi^{\rm usf}$ \\
                                & [\AA] & [GPa]    & [GPa]    & [GPa]    & [eV/\AA$^2$] \\ \hline\hline
        MTP & 3.172 & 551.4 & 200.0 & 139.9 & 1.17\\ \hline
        DFT & 3.168 & 545.3 & 211.5 & 145.5 & 1.11 \\ \hline
    \end{tabular}
    \caption{Comparison of the elastic properties and the 1/4\,[111] unstable stacking fault energy of the MTP for W to DFT.
    The DFT values of the elastic properties are taken from \cite{hodapp_operando_2020}, the DFT value of the stacking fault energy is taken from \cite{xu_frank-read_2020}}
    \label{tab:MTP_errors}
\end{table}

Next, we benchmark the MTP on properties of 1/2\,$\<$111$\>$ screw dislocations by comparing our "AutoPot MTP" to the MTP of \citet{hodapp_operando_2020} that has been trained on configurations that have been extracted on-the-fly while running a nudged elastic band calculation.
The MTP of \citet{hodapp_operando_2020} contains 22 configurations with 135 atoms each, comprising bulk neighborhoods, and atomic neighborhoods within the dislocation core.
Training a MLIP on all relevant neighborhoods appearing in a simulation is the most rigorous way to make the simulation using that MLIP as close as possible to a large-scale DFT simulation.
Therefore, we denote the MTP of \citet{hodapp_operando_2020} as the \emph{reference MTP} in the following.

\begin{figure}[hbt]
    \begin{minipage}{0.5\textwidth}
        \centering
        (a)
    \end{minipage}\hfill
    \begin{minipage}{0.5\textwidth}
        \centering
        (b)
    \end{minipage}\\[1em]
    \begin{minipage}{0.5\textwidth}
        \centering
        \includegraphics[width=0.67\textwidth]{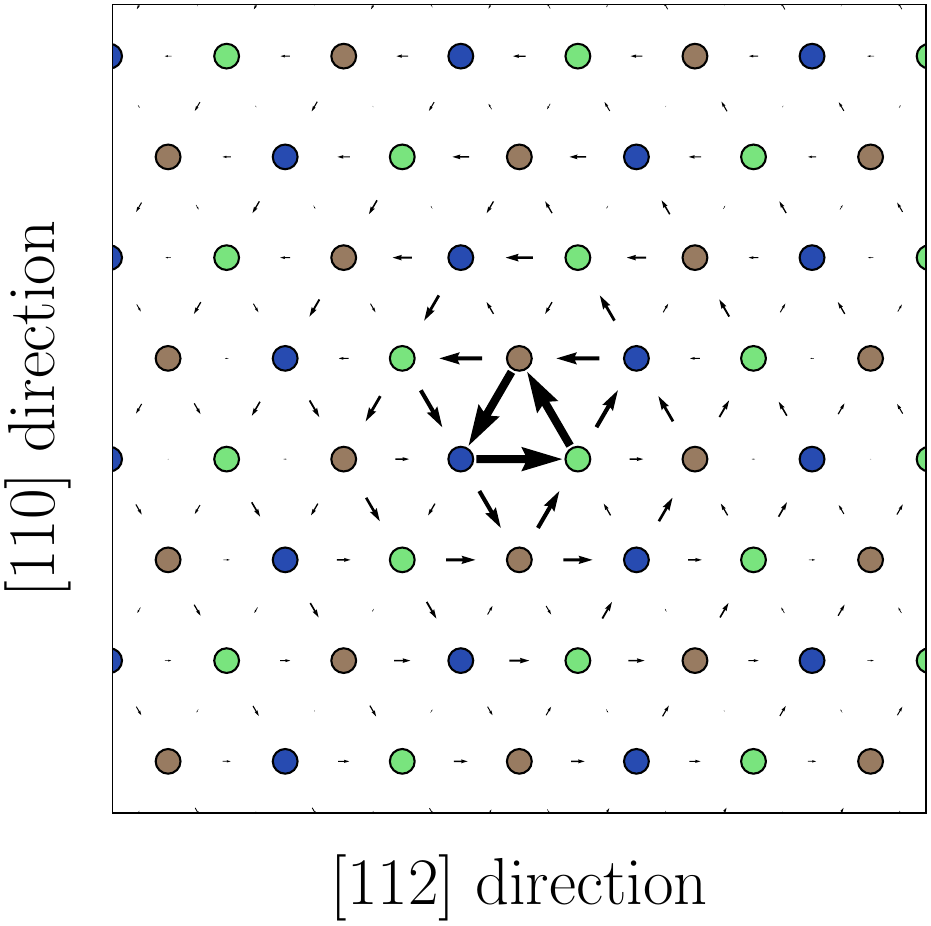}
    \end{minipage}\hfill
    \begin{minipage}{0.5\textwidth}
        \centering
        \includegraphics[width=0.9\textwidth]{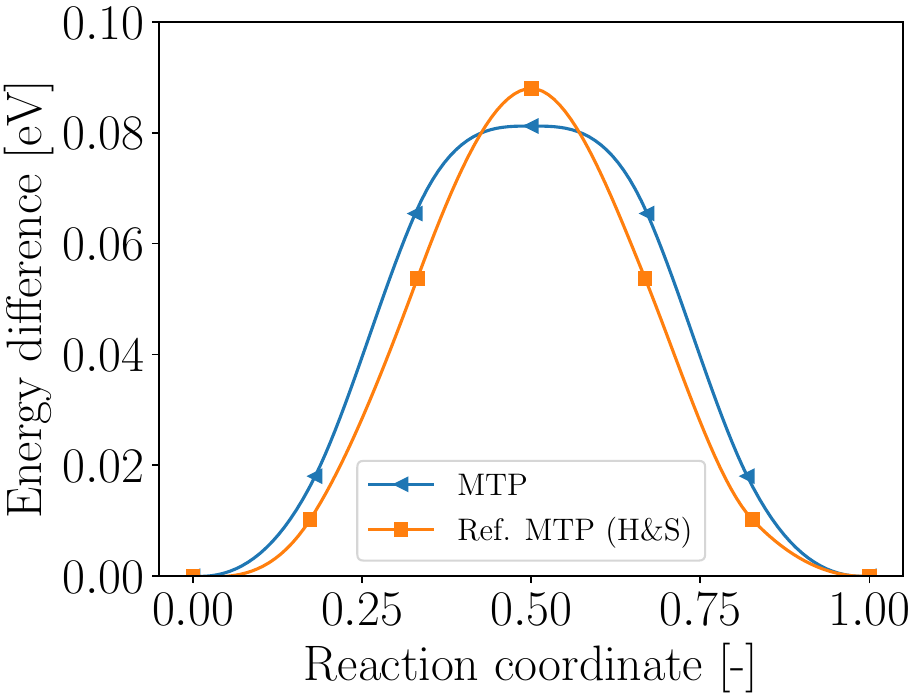}
    \end{minipage}
    \caption{(a) Differential displacement plot \cite{vitek_theory_1974} of the relaxed dislocation core using the MTP for W.
    (b) Minimum energy path of the MTP for W and of the reference MTP from \citet{hodapp_operando_2020} that has been trained on configurations containing the dislocation cores that appeared during the NEB calculations}
    \label{fig:dislocation_core_and_energy_curve}
\end{figure}

Upon relaxing the dislocation, we inspect the differential displacements around the dislocation core.
The relaxed core structure is non-polarized (Figure \ref{fig:dislocation_core_and_energy_curve} (a)) in agreement with the reference MTP and DFT (cf. \cite{hodapp_operando_2020}, Figure 3).
We then compute the energy difference between the relaxed configuration and the initial configuration in which the dislocation has been inserted using the elastic displacement field of a screw dislocation.
The MTP result (-0.158\,eV per Burgers vector) deviates only by a few percent from the reference MTP result (-0.166\,eV per Burgers vector).

We finally calculate the minimum energy path (MEP) of the screw dislocation while crossing the Peierls valley.
Again, the MEP of our new MTP, constructed using AutoPot, agrees very well with the MEP from the reference MTP;
the difference in the Peierls barrier of our MTP (0.081\,eV) is within a few percent from the reference MTP (0.088\,eV).
This is remarkable given the small size of the training set containing only 123 configurations, thanks to active learning selecting only the most distinct configuration to be used for training.

\subsection{MTP for Mo-Ta random alloys}

We compose the training set for Mo-Ta of the same types of configurations as for the W potential, namely, bulk configurations with 54 atoms per supercell, and configurations containing a 1/4\,[111] stacking fault with 72 atoms per supercell.
For each configuration type, we create a set of training candidates with 30\,000 samples using the lattice constant for the equiatomic alloy.
The alloy composition of each sample is drawn from a Dirichlet distribution over the entire composition space.
A Cauchy strain is applied to each sample which is extracted from a uniform distribution with the minimum and maximum strain components set to $\pm$\,3\,\%.
The atomic positions of each sample are randomly perturbed using a Gaussian distribution with a standard deviation of 0.02\,\AA.

Our initial training set contains 30 configurations (5 bulk and 5 stacking fault configurations at 100\,\%\,Mo, equiatomic composition, and 100\,\%\,Ta).
Pre-selection of training configurations from the candidate set is performed in the same way as for the W potential.
To test whether AutoPot is able to handle sampling from many MD trajectories in parallel, we run MD simulations on \emph{each} of the configurations that is contained in the training set after the pre-selection phase.

The evolution of the training set size as a function of the selection iteration is shown in Figure \ref{fig:selected_cfgs}.
The total number of configurations selected by AutoPot is 389.
Most of the configurations (287) are added during the second and last iteration of the pre-selection phase (when $\gamma_{\rm threshold} = 100\,000$, and $\gamma_{\rm threshold} = 2$).
Still a considerable amount of 102 configurations is selected during the MD simulations.
This can be attributed to the large number of MD simulations (287) but also to chemistry effects altering the dynamics of atoms, which is difficult to capture a priori before running a simulation (cf. \cite{moitzi_ab_2024}).

The MTP training errors are given in Table \ref{tab:MTP_errors}.
All errors are well within the tolerances that are known to be sufficient for MTPs being able to predict elastic and defect properties in random alloys with an accuracy close to that of DFT, i.e., of the order of 1\,meV for per-atom energies, of the order of 10\,meV/{\AA} for forces, and of the order of 10$^{-1}$\,GPa for stresses (cf., e.g, \cite{hodapp_machine-learning_2021,novikov_ai-accelerated_2022,moitzi_ab_2024}).

The MTP is then benchmarked on various properties relevant for studying dislocation plasticity, namely, the lattice constant, elastic constants, and the 1/4\,[111] unstable stacking fault energy;
the simulation details for all considered properties are given in Appendix \ref{sec:sim_details}.
Overall, the agreement with DFT over the entire composition space is very good, as shown in Table \ref{tab:MTP_errors}.
The MTP error on all quantities is less than 10\,\%, comparable to other accurate MLIPs for random alloys (e.g., \cite{li_complex_2020,hodapp_exact_2025,spitaler_ab-initio_2025}).

Overall, our results confirm the excellent accuracy of previous MTPs for alloys across the entire composition space.
However, those previous MTPs have been tediously constructed by actively sampling the training configurations from sequential MD simulations using semi-automated workflows (cf. \cite{hodapp_machine-learning_2021}).
With AutoPot, constructing highly accurate MTPs that are stable over a wide configuration and chemical space is now possible fully automatically within a couple of hours, and with minimal input required by the user.

\begin{figure}[hbt]
    \centering
    \begin{minipage}{0.5\textwidth}
        \centering
        (a)
    \end{minipage}\hfill
    \begin{minipage}{0.5\textwidth}
        \centering
        (c)
    \end{minipage}\\[1em]
    \begin{minipage}{0.5\textwidth}
        \centering
        \includegraphics[width=0.9\textwidth]{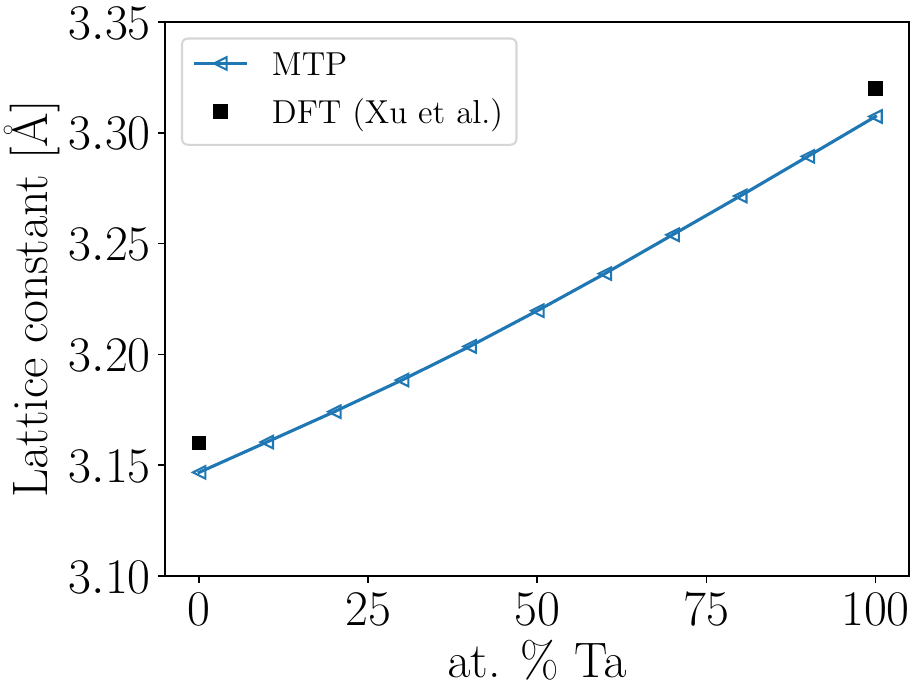}
    \end{minipage}\hfill
    \begin{minipage}{0.5\textwidth}
        \centering
        \includegraphics[width=0.9\textwidth]{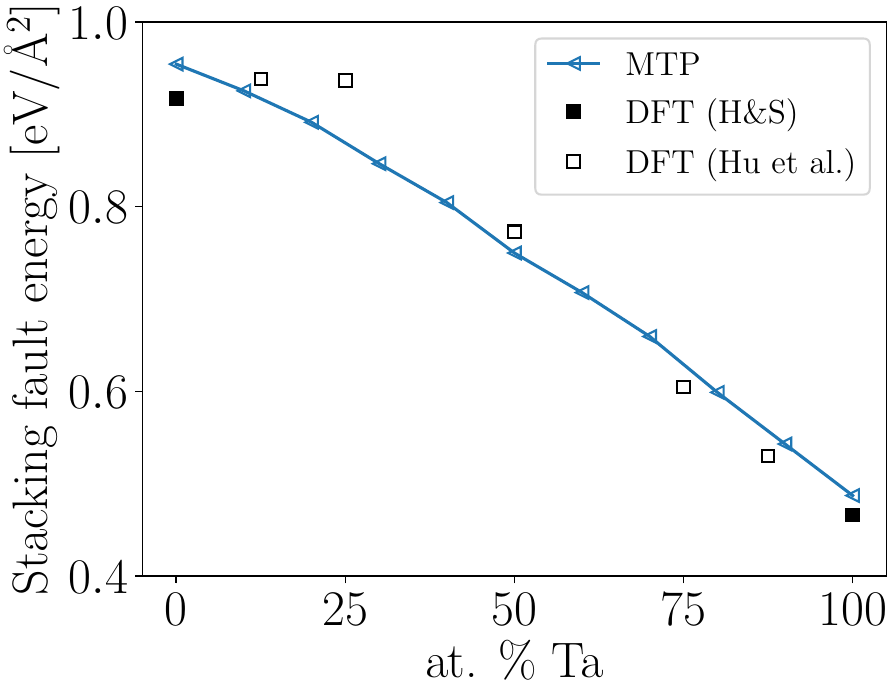}
    \end{minipage}\\[1em]
    \begin{minipage}{0.7\textwidth}
        \centering
        (b)
    \end{minipage}\\[1em]
    \begin{minipage}{0.75\textwidth}
        \centering
        \includegraphics[width=0.9\textwidth]{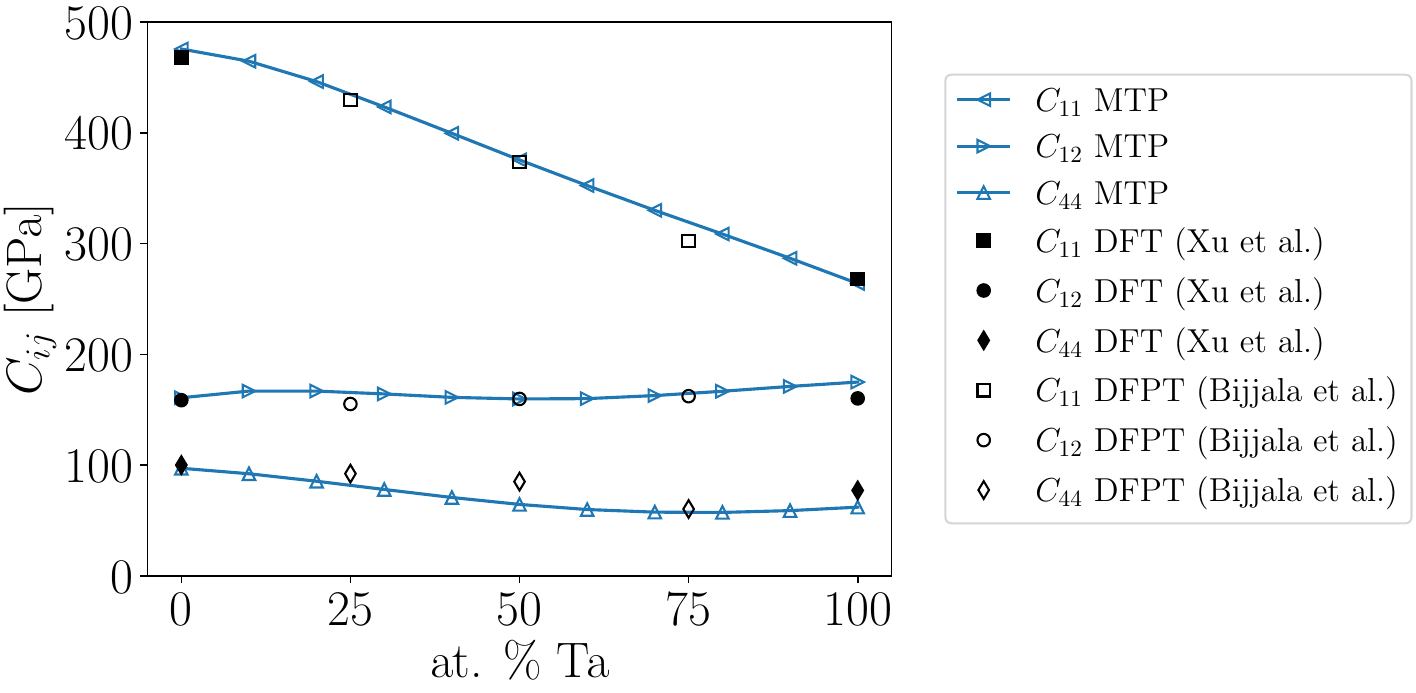}
    \end{minipage}
    \caption{Comparison of the elastic properties and the 1/4\,[111] unstable stacking fault energy of the MTP for Mo-Ta to DFT.
    The DFT values are taken from \citet{xu_frank-read_2020}, \citet{hu_screening_2021}, \citet{hodapp_machine-learning_2021}, and \citet{bijjala_elastic_2025}}
    \label{fig:MoTa_properties}
\end{figure}

\section{Concluding remarks}

In the present work, we have introduced AutoPot, a software for constructing machine-learning potentials (MLIPs) in an automated fashion.
MLIPs aim at bringing quantitative accuracy of quantum mechanics to the atomic scale, but reaching their full potential is still hampered by constructing proper training data, in particular for multicomponent systems, or systems with complex quantum-mechanical interactions, such as magnetism.
Advanced learning algorithms like active learning or fine-tuning can substantially reduce the amount of training data and improve the accuracy but require sophisticated protocols to streamline the construction of MLIP.
AutoPot alleviates designing such protocols using the BlackDynamite framework that manages parametric studies in a highly automated and parallelized fashion, supported by the Motoko workflow manager that manages interactions between the studies.

One of the key advantages of basing AutoPot on BlackDynamite is the possibility of allowing tasks to be implemented as Python scripts without the need of writing parsers.
As such it can easily be extended to include other MLIP software, ab initio codes, etc.
It therefore provides the community with a flexible and versatile tool for streamlining the construction of MLIPs.
As such, it can also serve as a framework for benchmarking learning protocols for MLIPs, i.e., benchmarking the expressivity and transferability of training sets generated using different sampling algorithms with different MLIPs, and comparing the performance of different MLIPs trained on the so-constructed training sets.
This is, in our view, a critically understudied part of constructing MLIPs, as benchmarking MLIPs is mostly done by comparing their accuracy on pre-defined datasets.
The present version of AutoPot selects training configurations based on the D-optimality criterion.
However, many other, possibly more advanced, algorithms have been proposed for different MLIPs, such as sampling from uncertainty-driven simulations \cite{kulichenko_uncertainty-driven_2023,fletcher_autonomous_2025}, sampling from configurations with an optimized (minimal) number of atoms \cite{salzbrenner_developments_2023,poul_automated_2025}, re-calibration of uncertainties \cite{thomas-mitchell_calibration_2023,ho_flexible_2025}, or property-specific error metrics \cite{liu_learning_2024}, which could be integrated in future versions of AutoPot, and benchmarked on different MLIPs.

For users who would like to treat AutoPot as a black box without touching the default parameters and the workflow itself, we currently recommend to use it only for cases where the configuration space and the simulated temperature range are known;
for example, if a defect to be investigated fits into a DFT supercell with $\sim$\,100 atoms used for the training.
An accurate DFT setup is also mandatory, in particular when many species (e.g., in the case of high-entropy alloys) are involved where self-consistent iterations may struggle to converge.

While our first version of AutoPot is presented as a standalone program, our vision is to interface it with a molecular dynamics code, where training configurations can be constructed and submitted on-the-fly, thanks to the splitting of job submission and workflow orchestration.
Developing such an interface will be part of future work.

Finally, we remark that the methodology developed for AutoPot is not limited to MLIPs but could be applied to other learning tasks, such as actively learning continuum models from atomistic data (e.g., \cite{tian_data-driven_2024}).

\section{Data availability}

BlackDynamite, Motoko, and AutoPot are on gitlab \cite{noauthor_blackdynamite_2024,anciaux_motoko,hodapp_autopot}.

\section{Acknowledgments}

We thank Franco Moitzi and Daniil Khodachenko for their initial testing of AutoPot and helpful suggestions on improving the documentation.

MH gratefully acknowledges the financial support by the Austrian Federal Ministry for Labour and Economy and the National Foundation for Research, Technology and Development and the Christian Doppler Research Association.

\section*{Appendix}

\begin{appendices}

\counterwithin*{equation}{section}
\renewcommand\theequation{\thesection\arabic{equation}}

\renewcommand{\thefigure}{A\arabic{figure}}
\setcounter{figure}{0}

\section{Autopot orchestration diagram}

The orchestration diagram of the full workflow implemented in AutoPot is shown in Figure \ref{fig:full_workflow}.

\afterpage{%
\clearpage  
\begin{landscape}
\begin{figure}
    \centering
    \includegraphics[width=0.999\linewidth]{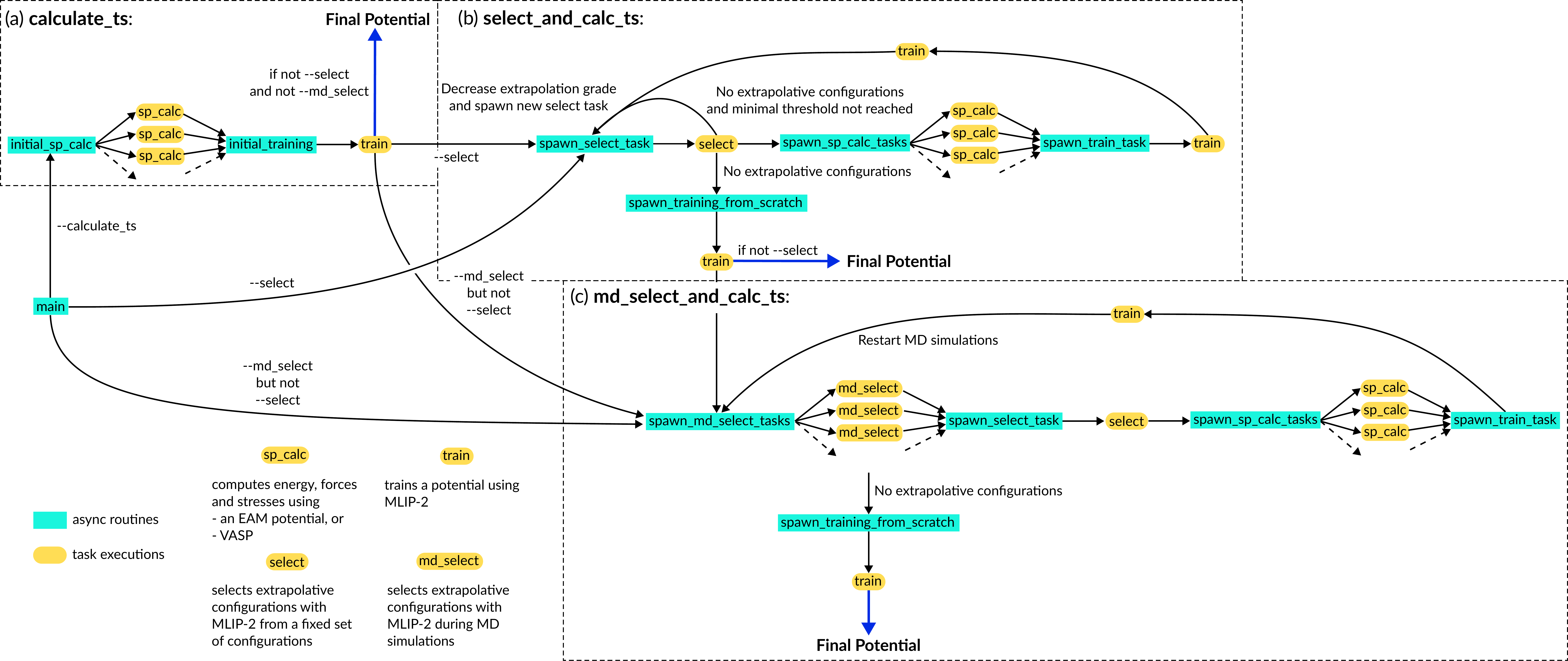}
    \vfill
    \caption{Full workflow implemented in AutoPot}
    \label{fig:full_workflow}
\end{figure}
\end{landscape}%
}  

\section{Simulation details}
\label{sec:sim_details}

Below, we give the simulation details of all the calculations from Section \ref{sec:production_mode}.
Setting up the simulations has been done in Python with the help of ASE \cite{hjorth_larsen_atomic_2017}, matscipy \cite{grigorev_matscipy_2024}, and MlipPyKit (\url{gitlab.com/mhodapp/mlippykit}).

\paragraph{Ground state}

To compute the material's ground state, we minimize the total energy of the primitive cell with respect to the lattice constant using the Nelder-Mead simplex algorithm as implemented in \texttt{SciPy}.
We consider a configuration as converged when the difference between the solutions for two subsequent iterations is less than $10^{-10}$\,\AA.
For computing the lattice constants of the alloys, we use a supercell of 2\,000 atoms with the Mo and Ta atoms randomly distributed corresponding to a given composition.

\paragraph{Elastic constants}

We compute the elastic constants by linearly regressing the stresses as a function of the strains around the ground state using the \texttt{fit\_elastic\_constants} function from matscipy.
For computing the elastic constants of the alloys, we again use a supercell of 2\,000 atoms with the Mo and Ta atoms randomly distributed corresponding to a given composition.

\paragraph{\texorpdfstring{$1/4[111]$}{1/4[111]} stacking fault energy}

We first create a rectangular supercell with the directions corresponding to the axes set to $[11\overline{2}]$, $[\overline{1}10]$, and $[111]$, respectively.
This configuration is denoted by $\{ \ur^i \}_{\rm bulk}$.
We now translate half of the atoms in $\{ \ur^i \}_{\rm bulk}$ by one Burgers vector in the $[111]$ direction to create the stacking fault;
in addition, we apply a shear displacement of half a Burgers vector to the cell vectors such that there will be only one stacking fault per supercell.
This configuration is denoted by $\{ \ur^i \}_{\rm sf}$.
The stacking fault energy is the difference between the total energy of both configurations divided by the area of the fault plane $A$
\[
    \Pi^{\rm sf} = \frac{\Pi(\{ \ur^i \}_{\rm sf}) - \Pi(\{ \ur^i \}_{\rm bulk})}{A}
    .
\]
For computing $\Pi^{\rm sf}$ of a random alloy, we use supercells of about 100\,000 atoms, which practically removes any spurious size effects (cf. \cite{moitzi_ab_2024}).

\paragraph{Dislocation core relaxation}

The dislocation is inserted in a cylindrical configuration with a radius of 35\,{\AA} times the magnitude of the Burgers vector.
We constrain the positions of the outermost atoms that have a radial distance to the boundary of two times the cut-off radius to the linear elastic solution of a screw dislocation.
This leaves about \,5\,000 free atoms that are allowed to relax.
Energy minimization is performed using the Fast Inertial Relaxation Engine \cite{bitzek_structural_2006}, as implemented in ASE.
We consider an energy minimization as converged when the maximum absolute force on an atom is less than $10^{-3}$\,eV/\AA.

\paragraph{NEB calculations}

The NEB calculations are performed using the same configuration as for the dislocation core relaxation.
We use the climbing NEB method of \citet{henkelman_improved_2000} with seven images.
The images are created by linearly interpolating between two configurations that differ in the position of the dislocation by one Peierls distance.

\end{appendices}

\printbibliography[heading=bibintoc]

\end{document}

%% file: style.tex
\newcommand{\sty}[1]{\boldsymbol{#1}}

\newcommand{\ubar}[1]{\mkern 0.5mu\underline{\mkern-0.5mu#1\mkern-0.5mu}\mkern 0.5mu}
\newcommand{\uubar}[1]{\ubar{\ubar{#1}}}

\let\epsilon\varepsilon
\let\mtheta\theta
\let\theta\vartheta

\let\rho\varrho

\let\phi\varphi
\let\Gamma\varGamma
\let\Delta\varDelta
\let\Theta\varTheta
\let\Lambda\varLambda
\let\Xi\varXi
\let\Pi\varPi
\let\Sigma\varSigma
\let\Upsilon\varUpsilon
\let\Phi\varPhi
\let\Psi\varPsi

\let\Omega\varOmega

\newcommand{\bmk}{\sty{k}}

\newcommand{\clE}{\mathcal{E}}

\newcommand{\sT}{\mathsf{T}} 

\newcommand{\scL}{\mathscr{L}}

\newcommand{\scN}{\mathscr{N}}

\newcommand{\scT}{\mathscr{T}}

\newcommand{\uc}{\ubar{c}}

\newcommand{\uf}{\ubar{f}}

\newcommand{\ur}{\ubar{r}}

\newcommand{\umtheta}{\ubar{\mtheta}}

\newcommand{\uuA}{\uubar{A}}

\newcommand{\uusigma}{\uubar{\sigma}}

\newcommand{\<}{\langle}
\renewcommand{\>}{\rangle}

\newcolumntype{x}[1]{>{\centering\arraybackslash\hspace{0pt}}p{#1}}